\documentclass[twocolumn,aps,prb,amsmath,floatfix]{revtex4}
\usepackage{graphicx}
\begin{document}
\title{Exact Diagonalization Dynamical Mean Field Theory for 
Multi-Band Materials:\\ Effect of Coulomb correlations on the 
Fermi surface of Na$_{0.3}$CoO$_2$ }
\author{C.A. Perroni,$^1$ H. Ishida,$^2$ and A.~Liebsch$^1$} 
\affiliation{$^1$Institut f\"ur Festk\"orperforschung, 
             Forschungszentrum J\"ulich, 
             52425 J\"ulich, Germany\\
             $^2$CREST, JST, and College of Humanities and Sciences, 
             Nihon University, Sakura-josui, Tokyo 156, Japan}
\begin{abstract} 
Dynamical mean field theory combined with finite-temperature
exact diagonalization is shown to be a suitable method to study
local Coulomb correlations in realistic multi-band materials. 
By making use of the sparseness of the impurity Hamiltonian, 
exact eigenstates can be evaluated for significantly larger 
clusters than in schemes based on full diagonalization. Since
finite-size effects are greatly reduced this approach allows 
the study of three-band systems down to very low temperatures, 
for strong local Coulomb interactions and full Hund exchange. 
It is also shown that exact diagonalization yields smooth subband 
quasi-particle spectra and self-energies at real frequencies.
As a first application the correlation induced charge transfer 
between $t_{2g}$ bands in Na$_{0.3}$CoO$_2$ is investigated. 
For both Hund and Ising exchange the small $e'_g$ Fermi surface 
hole pockets are found to be slightly enlarged compared to the 
non-interacting limit, in agreement with previous Quantum Monte 
Carlo dynamical mean field calculations for Ising exchange, but 
in conflict with photoemission data.
    
\end{abstract}
\maketitle

\section{Introduction} 
Dynamical mean field theory (DMFT) \cite{rmp} has been used 
successfully during recent years to describe the electronic 
properties of a variety of strongly correlated materials. 
\cite{reviews} The hallmark of these systems are their complex
lattice geometries, giving rise to intricate single-particle 
properties, accompanied by complex many-electron interactions 
in partially filled, nearly localized atomic orbitals. These 
characteristics lead to a wealth of physical phenomena such as 
metal insulator transitions, exotic magnetic structures and 
unconventional superconducting phases.\cite{imada} The virtue of 
DMFT is that it treats single-electron and many-electron features 
on the same footing. The key conceptual advance of this approach
is the mapping of the lattice problem onto an effective impurity
problem which is solved numerically exactly. The local self-energy 
is then determined via a self-consistency procedure.     

The many-body impurity problem can be solved, in the case of 
single-band systems, by using techniques such as
numerical renormalization group \cite{nrg} (NRG),
exact diagonalization\cite{ed} (ED),
quantum Monte Carlo\cite{qmc} (QMC), 
or other schemes.\cite{rmp} 
For computational reasons dynamical correlations in realistic 
multi-band materials 
have so far been investigated mainly within QMC\cite{reviews},
iterated perturbation theory\cite{ipt} (IPT) and the 
fluctuation exchange method\cite{flex} (FLEX).
The versatility of the QMC approach is made feasible by allowing only 
for Ising-like exchange interactions to avoid serious sign problems 
at low temperatures.\cite{held} Extensions of QMC including full 
Hund's exchange are presently limited to $T=0$ \cite{arita} and 
rather high temperatures ($T>1500$~K).\cite{koga3} 
Spin-flip interactions at low finite temperatures can be taken 
into account in two recently developed schemes, namely, the  
continuous-time QMC method \cite{rubtsov,werner} and the combination 
of the Hirsch-Fye algorithm with a perturbation series expansion. 
\cite{sakai}    

The aim of this work is to demonstrate that multi-band ED/DMFT is a 
highly useful scheme for the investigation of Coulomb correlations
in realistic materials. In the past, applications of ED/DMFT were 
limited to one-band and two-band systems because of the extremely 
rapid increase of the Hilbert space when the size $n_s$ of the cluster 
used to simulate the solid is enlarged. \cite{koga1,prl05,biermann}
For instance, ED with two impurity orbitals, each coupled to three 
bath levels ($n_s=8$), requires diagonalization of matrices with 
dimension up to 4900 which roughly represents the time and storage 
limit of what is computationally meaningful.\cite{al+costi} 
We show here that a considerable simplification of this task 
can be achieved by exploiting the extreme sparseness of the 
Hamiltonian matrices and focusing on the limited number of 
excited states relevant at low temperatures. As a result
the cluster size $n_s=12$ can be treated at about the same 
computational cost as full diagonalization for $n_s=8$.
This improvement allows for the first time the application
of ED/DMFT to realistic three-band systems.  

In this work we study the intercalated layer compound Na$_x$CoO$_2$ 
which exhibits an unusual range of electronic properties as a 
function of Na doping:  unconventional superconductivity, 
magnetic and charge ordering, and metal insulator transition. 
Here we focus on one of the most puzzling and controversial 
aspects, namely, the topology of the Fermi surface at $x=0.3$, 
the Na concentration at which the system becomes superconducting
when additionally doped with water.\cite{takada} 
As a result of the octahedral crystal field, the Co $3d$ bands 
are split into $t_{2g}$ and $e_g$ subbands separated by a finite
energy gap. With Na doping the filling of the $t_{2g}$ 
bands can be continuously tuned between $n=5$ and $n=6$.   
The rhombohedral symmetry at Co sites yields a further splitting
of $t_{2g}$ orbitals into an $a_g$ and two doubly degenerate 
$e'_{g}$ orbitals.   
According to density functional theory within the local density 
approximation (LDA) the Fermi surface consists of a large $a_g$
hole surface centered at $\rm \Gamma$ and six small $e'_g$ hole 
pockets along the $\rm\Gamma K$ directions.\cite{singh} At $x=0.3$ 
the 0.7 holes in the $t_{2g}$ band consist of about 0.4 $a_g$ holes 
and 0.3 $e'_g$ holes. Thus, both types of states should be important
for a variety of electronic and magnetic properties. 
     
In striking contrast to these LDA predictions, angle-resolved 
photoemission spectroscopy (ARPES) experiments reveal only the
large $a_g$ Fermi surface, the $e'_g$ bands being completely 
filled.\cite{arpes}   
To resolve this discrepancy several possible 
explanations have been discussed in the literature. 
Since the onsite Coulomb energy $U$ among Co $3d$ electrons 
is about twice the width of the $t_{2g}$ bands a correlation 
induced inter-orbital charge transfer among $t_{2g}$ states 
could in principle modify the shape of the Fermi surface. 
Dynamical correlations evaluated within multi-band QMC/DMFT 
for a realistic single-particle Hamiltonian predict 
slightly enlarged $e'_g$ Fermi surface pockets.\cite{ishida} 
The same trend was found in recent QMC/DMFT calculations.
\cite{lechermann,marianetti}
On the other hand, accounting approximately for band narrowing 
via the Gutzwiller approach, filled $e'_g$ bands were obtained 
in the strong-coupling $U\rightarrow\infty$ limit,\cite{gutzwiller}
but these calculations have not yet been extended to realistic 
values of $U$ and finite exchange integrals $J$.

The role of the intercalated Na atoms has also been studied.
Within LDA, disorder in the Na layer was shown to produce 
potential variations that can localize the $e'_g$ Fermi 
surface pockets, at least for large doping.\cite{singh+deepa} 
A recent LDA+QMC/DMFT study of Na disorder also showed that the 
pockets are likely to disappear at large $x$, but to remain
stable near $x=0.3$.\cite{marianetti}

Although the QMC/DMFT results\cite{ishida,lechermann,marianetti}  
so far represent the 
most accurate analysis of correlation induced modifications of 
the Fermi surface of Na$_{0.3}$CoO$_2$, they are limited to 
Ising exchange because of the QMC sign problems alluded to above. 
Since the $e'_g$ bands reach only about 100~meV above the Fermi 
level -- an energy much smaller than typical exchange integrals --  
there arises the question to what extent the inclusion 
of full Hund's exchange might provide an explanation of the 
observed ARPES data.

To address this issue we have applied 
ED/DMFT to Na$_{0.3}$CoO$_2$. The important result 
of the present work is that dynamical on-site correlations
within the Co $t_{2g}$ manifold for realistic values of 
Coulomb and exchange energies give rise to a transfer of 
electronic charge from the $e'_g$ bands to the $a_g$ band. 
For Hund's exchange and $T=10$~meV the number of $a_g$ holes
is decreased to $0.351$ compared to the LDA value 
$0.40$, while the number of $e'_g$ holes is increased to
$0.349$ from the LDA value $0.30$. Nearly identical 
values are found for Ising exchange: $0.345$  ($a_g$) and 
$0.355$ ($e'_g$). These results are qualitatively consistent 
with previous QMC/DMFT calculations for Ising exchange
at $T=30\ldots 60$~meV. Local Coulomb correlations can 
therefore be ruled out as a possible explanation of the 
absence of the $e'_g$ hole pockets in the ARPES data. 

It remains to be investigated experimentally and theoretically 
to what extent the photoemission data are influenced by surface 
effects which have played a crucial role also in other transition
metal oxides, such as Sr$_2$RuO$_4$ and Sr$_x$Ca$_{1-x}$VO$_3$.   
Also, we note that the present ARPES results do not seem to be 
consistent with recent Shubnikov-de Haas measurments 
of the Fermi surface of  Na$_{0.3}$CoO$_2$.\cite{balicas}

The present work establishes multi-band ED/DMFT as a useful 
scheme for describing strongly correlated materials. 
In this sense ED/DMFT can now be regarded as complementary 
to multi-band QMC/DMFT. In comparison to standard Hirsch-Fye
QMC treatments, important advantages of ED/DMFT are that 
spin-flip and pair-exchange interactions are fully taken 
into account, large Coulomb energies pose no computational 
problems, and temperatures as low as 5 to 10 meV are 
readily accessible. 

The outline of this paper is as follows. In Section II we
present the main ingredients of the multi-band ED/DMFT approach.
We also argue that it is possible within this scheme to 
evaluate continuous spectral functions and self-energies at 
real frequencies for the extended material. 
Section III presents the application to Na$_{0.3}$CoO$_2$ 
where we focus on the correlation induced charge transfer 
among the partially filled $t_{2g}$ bands. Summary and Outlook 
are given in Section IV.     

\section{Theory}
\subsection{Multi-band ED/DMFT}

Let us consider a material whose single-particle properties are
characterized by a Hamiltonian $H({\bf k})$. In the case of 
partially occupied $t_{2g}$ bands of a transition metal oxide
such as Na$_x$CoO$_2$, $H({\bf k})$ is a $3\times3$ matrix 
whose elements account for direct interactions between
$t_{2g}$ orbitals as well as indirect interactions via 
neighboring Oxygen ions. In addition we consider on-site 
Coulomb interactions which are comparable to or larger than 
the $t_{2g}$ band width. The purpose of single-site DMFT is 
to derive a local self-energy $\Sigma(\omega)$ which describes 
the modification of the single-particle bands caused by Coulomb 
interactions. The local lattice Green's function is then given 
by the expression   
\begin{equation}
   G_{\alpha\beta} (i\omega_n)  = \sum_{\bf k} \Big(
   i\omega_n + \mu - H({\bf k})  - \Sigma(i\omega_n)
          \Big)^{-1}_{\alpha\beta}
                                                   \label{G}
\end{equation}
where $\omega_n=(2n+1)\pi/k_B T$ are Matsubara frequencies, 
$\mu$ is the chemical potential, 
and $\alpha,\beta = d_{xy}, d_{xz}, d_{yz}$ denotes the 
$t_{2g}$ orbital basis.
We consider paramagnetic systems and omit the spin index of 
Green's functions and self-energies for convenience. 

In the case of a two-dimensional hexagonal lattice, local 
quantities such as $G_{\alpha\beta} (i\omega_n)$ have only 
two independent elements given by $G_{11}=G_{22}=G_{33}$ and
$G_{12}=G_{13}=G_{23}$. It is therefore convenient to go over 
to the the $a_g$, $e'_g$ basis, where
$a_g=(d_{xy}+d_{xz}+d_{yz})/\sqrt{3}$,
$e'_{g1}=(d_{xy}+d_{xz}-2d_{yz})/\sqrt{6}$ and  
$e'_{g2}=(d_{xy}-d_{xz})/\sqrt{2}$.   
Within this basis $G$ becomes diagonal, with elements
\begin{eqnarray}
 G_{m=1} &\equiv& G_{a_g} \ = \ G_{11}+2G_{12}  \\  
 G_{m=2,3} &\equiv& G_{e'_g} \ = \ G_{11}-G_{12}. 
\end{eqnarray}
The reverse transformation is $G_{ii}=(G_{a_g}+2G_{e'_g})/3$ 
and $G_{ij}=(G_{a_g}- G_{e'_g})/3$. Except for the Brillouin
Zone integral in Eq.~(\ref{G}) we perform all subsequent 
calculations in the diagonal $a_g$, $e'_g$ basis denoted by 
$m=1\ldots3$. Analogous transformations hold for other 
lattice structures, such as square two-dimensional or 
cubic three-dimensional systems.

For the purpose of the quantum impurity calculation it is
necessary to first remove the self-energy from the central site. 
This step yields the impurity Green's function     
\begin{equation}
G_{0,m}(i\omega_n)=[G_m(i\omega_n)^{-1}+\Sigma_m(i\omega_n)]^{-1}.
                                                      \label{G0}
\end{equation} 
Note that, since $G_{m}$, $G_{0,m}$ and $\Sigma_{m}$ characterize
the extended solid, they have smooth, continuous spectra at real 
frequencies.  

Within ED/DMFT\cite{ed} the lattice impurity Green's function 
$G_{0,m}$ is approximated via an Anderson impurity model for a 
cluster consisting of impurity levels $\varepsilon_{m=1\ldots3}$ 
and bath levels $\varepsilon_k={4\ldots n_s}$\, coupled via 
hopping matrix elements $V_{mk}$. Thus, 
\begin{equation}
     G_{0,m}(i\omega_n) \approx  G_{0,m}^{cl}(i\omega_n)   \label{G00}
\end{equation} 
where
\begin{equation}
 G_{0,m}^{cl}(i\omega_n) = \Big(i\omega_n + \mu - 
         \varepsilon_m - \sum_{k=4}^{n_s} 
         \frac{ \vert V_{mk} \vert^2 }
         {i\omega_n - \varepsilon_{k}}\Big)^{-1}  .
                                                       \label{G0cl}
\end{equation} 
Here, $n_s=12$ is the cluster size used in the calculations 
discussed below. Since $G_{0,m}^{cl}$ is diagonal in orbital 
indices, each impurity level couples to its own bath containing 
three levels. Evidently, in contrast to Im\,$G_{0,m}$, at real 
$\omega$ Im\,$G_{0,m}^{cl}$ is discrete, with the number of 
poles determined by the cluster size.

The impurity Hamiltonian for the cluster is defined as:
\begin{eqnarray}
   H &=& \sum_{m\sigma}  (\varepsilon_{m}-\mu) n_{m\sigma} \  +
         \sum_{k\sigma} \varepsilon_{k} n_{k\sigma}
                                                     \nonumber\\ 
     && + \sum_{mk\sigma} V_{mk}
            [ c_{m\sigma}^+ c_{k\sigma} + {\rm H.c.} ]   \nonumber\\
  && + \sum_{m} U n_{m\uparrow} n_{m\downarrow}  + \!\!\!\!
  \sum_{m< m'\sigma\sigma'}\!\!\!\! (U'-J\delta_{\sigma\sigma'}) 
                 n_{m\sigma} n_{m'\sigma'}                \nonumber\\
  && -\sum_{mm'} J'[\,c_{m\uparrow}^+ c_{m\downarrow}
            c_{m'\downarrow}^+ c_{m'\uparrow}   
           + c_{m\uparrow}^+ c_{m\downarrow}^+    
            c_{m'\uparrow} c_{m'\downarrow}] 
                                               \label{Hcl}
\end{eqnarray}
where $c_{m\sigma}^{(+)}$ are annihilation (creation) operators 
for electrons in impurity level $m\le3$ with spin $\sigma$ and  
$n_{m\sigma}=c_{m\sigma}^+ c_{m\sigma}$, with a similar notation
for the bath levels $k=4\ldots n_s$. 
H.c.~denotes Hermitian conjugate terms.
The atomic part of the Hamiltonian is identical with the atomic
part of the Hamiltonian of the extended material.  
The intra- and inter-orbital impurity Coulomb energies are 
denoted by $U$ and $U'$. The exchange integral is $J$.
Because of rotational invariance it obeys the relation $U'=U-2J$. 
Spin-flip and pair-exchange terms are denoted explicitly by $J'$. 
In the case of isotropic Hund exchange, one has $J'=J$. In the 
case of Ising-like exchange these terms are neglected, so that $J'=0$. 

Within the diagonal $a_g$, $e'_g$ basis, the cluster Green's 
function has the spectral representation:\cite{rmp,capone}
\begin{eqnarray}
 G^{cl}_{m}(i\omega_n)&=&\frac{1}{Z} \sum_{\nu\mu} 
  \frac{\vert\langle \mu|c_{m\sigma}^+ |\nu \rangle \vert^2}
                    {E_\nu - E_\mu - i\omega_n}
   [e^{-\beta E_\nu} + e^{-\beta E_\mu}] \nonumber\\
 &=&\frac{1}{Z} \sum_{\nu} e^{-\beta E_\nu}  
    [G_{m\sigma}^{\nu+}(i\omega_n) + G_{m\sigma}^{\nu-}(i\omega_n)]\
                                                       \label{Gcl}
\end{eqnarray}
where $E_\nu$ and $|\nu\rangle$ are eigenvalues and eigenvectors 
of the impurity Hamiltonian and $Z=\sum_\nu {\rm exp}(-\beta E_\nu)$
is the partition function.
The excited state Green's functions are given by
\begin{eqnarray}
 G_{m\sigma}^{\nu+}(i\omega_n) &=& \sum_{\mu}
  \frac{\vert\langle \mu|c_{m\sigma}^+|\nu \rangle \vert^2}
                    {E_\nu - E_\mu - i\omega_n} ,  \label{Gm+} \\ 
 G_{m\sigma}^{\nu-}(i\omega_n) &=& \sum_{\mu}
\frac{\vert\langle \mu|c_{m\sigma} |\nu \rangle \vert^2}
{E_\mu - E_\nu - i\omega_n} .
                                                \label{Gm-}
\end{eqnarray}

In analogy to Eq.~(\ref{G0}) the cluster self-energy is given 
by the expression:
\begin{equation} 
\Sigma^{cl}_m(i\omega_n) = G^{cl}_{0,m}(i\omega_n)^{-1} - 
                           G^{cl}_m(i\omega_n)^{-1}. 
                                                \label{sigcl}
\end{equation}
As will be discussed in more detail in Subsection C,
it is useful to regard the self-energy as the central quantity 
determined via DMFT. The key physical assumption within the ED 
approach is then that the cluster self-energy is an adequate 
representation for the self-energy of the extended solid, i.e., 
\begin{equation} 
      \Sigma_m     (i\omega_n) \approx 
      \Sigma_m^{cl}(i\omega_n)\, .
\end{equation}
From Eq.~(\ref{G0}) it now follows that
\begin{equation} 
      G_m(i\omega_n) \approx G_m^{cl}(i\omega_n) .
\end{equation}
The approximate equalities here are important since they ensure 
that, as in the case of $G_{0,m}$, at real $\omega$ two alternative 
representations exist: a continuous one for the lattice Green's 
function $G_m$ and self-energy $\Sigma_m$, and a discrete one for 
the cluster Green's function $G_m^{cl}$ and self-energy $\Sigma_m^{cl}$.
(See Fig.~13 of Ref.~1 for a nice illustration of the fact 
that continuous and discrete real-$\omega$ spectra can nearly
coincide at Matsubara frequencies.)

Transforming from the $a_g$, $e'_g$ basis to the $t_{2g}$ 
basis via $\Sigma_{ii}=\Sigma_{a_g}+2\Sigma_{e'_g}$,
 $\Sigma_{i\ne j}=\Sigma_{a_g}-\Sigma_{e'_g}$, we insert this 
self-energy into the solid Green's function Eq.~(\ref{G}). 
The iteration cycle then has the schematic form: 
\begin{equation}
\Sigma   \rightarrow G      \rightarrow G_0  \approx
 G_0^{cl} \rightarrow G^{cl} \rightarrow \Sigma^{cl}
 \approx \Sigma .
\end{equation}
This procedure is repeated until a self-consistent solution 
is found for a given chemical potential. This potential is then 
varied until the calculated total occupancy of the $t_{2g}$
bands agrees with the desired occupancy.  

In practice, we start from the non-interacting case, where the 
cluster levels $\varepsilon_m$ and $\varepsilon_k$ and hopping 
matrix elements $V_{mk}$ are found for 
$G_{0,m}(i\omega_n) = G_{m}(i\omega_n)$ with $\Sigma_m=0$.
These cluster parameters are then used to begin the iteration
at finite Coulomb and exchange energies for a given temperature.
Finally, the spectral distributions are given by 
\begin{equation} 
 A_m(\omega)=-\frac{1}{\pi}{\rm Im}\,G_m(\omega) \ . \label{A}
\end{equation}
The transformation of $G_{m}(i\omega_n)$ and $\Sigma_{m}(i\omega_n)$ 
to real frequencies will be discussed in the following subsection. 

At low temperatures, the Boltzmann factors in Eq.~(\ref{Gcl}) 
ensure that only a small number of excited states $|\nu\rangle$ 
are needed. Because of the sparseness of the impurity 
Hamiltonian matrix, these states can be evaluated exactly by 
using the Arnoldi algorithm.\cite{arnoldi} For a given excited 
state the Green's functions $G_{m\sigma}^{\nu\pm}(i\omega_n)$ can 
then be calculated very accurately using the Lanczos procedure, by 
starting from the vectors $c_{m\sigma}^+ |\nu \rangle$ 
and $c_{m\sigma}|\nu \rangle$, respectively:
\begin{eqnarray}
 G_{m\sigma}^{\nu+}(i\omega_n) &=&
 \frac{     \langle\nu|c_{m\sigma}c_{m\sigma}^+|\nu \rangle        }
          {a_{0+} - i\omega_n + \frac{b_{1+}^2}
          {a_{1+} - i\omega_n + {b_{2+}^2}/\ldots}}, \label{L+}\\
 G_{m\sigma}^{\nu-}(i\omega_n) &=&
 \frac{     \langle\nu|c_{m\sigma}^+c_{m\sigma}|\nu \rangle        }
          { a_{0-} - i\omega_n + \frac{b_{1-}^2}
          { a_{1-} - i\omega_n + {b_{2-}^2}/\ldots}}. \label{L-}
\end{eqnarray}

The reformulation of the cluster Green's function  
in Eq.~(\ref{Gcl}), (\ref{Gm+}), (\ref{Gm-}) is similar to 
the one used by Capone {\rm et al.} \cite{capone} 
in the single-band case, except that these authors calculate 
the excited states $|\nu \rangle$ by applying the Lanczos method.
This is faster than the exact solution if only few excited 
states are important. However, once extensive re-orthogonalization 
is required to obtain sufficiently accurate excited states, the 
Lanczos method becomes increasingly time-consuming so that the 
exact approach is preferable. 

The Arnoldi algorithm is very useful since 
it enormously reduces both storage and time requirements to 
carry out ED/DMFT calculations. For $n_s=12$ the largest
Hilbert space for $n_\uparrow = n_\downarrow =6$ has dimension
$N=853776$. Nevertheless, at most $M=23$ elements are finite 
in any given row of the Hamiltonian matrix. The size of the 
effective $N\times M$ matrix is therefore less than for full
diagonalization of the largest sector (4900) in the case $n_s=8$. 
Moreover, at temperatures of about $T=10$~meV fewer than 30 of 
the 169 possible $(n_\uparrow,n_\downarrow)$ configurations have 
lowest eigenvalues with Boltzmann factors larger than $10^{-5}$.  
Most of these sectors contribute only few excited states. Only in 
some $(n_\uparrow,n_\downarrow)$ sectors 20 to 40 eigenstates are 
important. Thus for a total of about 300 excited states the 
Green's functions $G^{\nu\pm}_{m\sigma}(i\omega_n)$ are calculated 
via the Lanczos procedure. Of course, this number increases at 
higher temperatures, and if Boltzmann factors smaller than 
$10^{-5}$ are included for higher precision.    
Finally, the Arnoldi algorithm requires as main computational 
step the matrix times vector operation $H u = v$ where $H$ has 
effective dimension $N\times M$, the vectors $u,v$ are of length 
$N$ and the internal sum extends only over the finite elements 
of $H$. Thus, in contrast to full diagonalization, this method 
is ideally suited for parallelization. Using 16 processors one 
iteration step takes less than 30 Min.

We have tested the above procedure on the two-band model 
consisting of half-filled subbands of different widths and found 
excellent agreement with previous results for $n_s=8$ based on 
full diagonalization.\cite{al+costi}      

\subsection{Quasi-particle spectra at real frequencies}

In the ED/DMFT equations given above we explicitly distinguish,
via the superscript $cl$, the cluster Green's functions and 
self-energy from the corresponding quantities of the extended
solid. This is done in order to emphasize that ED/DMFT is well 
suited for evaluating continuous solid quasi-particle spectra 
at real frequencies, as we now explain.  

The key point is that the solid spectra are to be derived from 
$G_m(\omega)$, Eq.~(\ref{G}), rather than $G^{cl}_m(\omega)$, 
Eq.~(\ref{Gcl}). Although in 
the self-consistent limit both functions agree at Matsubara 
frequencies within some uncertainty determined by the cluster
size, they differ fundamentally at real frequencies, with 
$G_m(\omega)$ exhibiting a smooth spectrum and $G^{cl}_m(\omega)$ 
consisting of $\delta$ functions. Within the spirit of 
finite-temperature ED/DMFT each iteration step involves the 
evaluation of the solid Green's function $G_{m}$ via Eq.~(\ref{G}) 
and the subsequent approximation of the associated impurity 
function $G_{0,m}$ in Eq.~(\ref{G0}) 
via the discrete cluster spectrum of $G_{0,m}^{cl}$, Eq.~(\ref{G0cl}). 
At a given temperature, the quality of this projection onto the set
of Matsubara frequencies depends strongly on the cluster size, i.e.,
on the number of poles at real $\omega$.

The task at the end of the iteration procedure, of transforming 
the lattice subband Green's functions $G_m(i\omega_n)$ from 
Matsubara frequencies to real $\omega$, is therefore similar 
to the one faced in QMC/DMFT. The difference is that the ED/DMFT 
data exhibit finite-size effects rather than statistical 
uncertainties. 

Since the ED/DMFT calculations are free of statistical effects,
a convenient extrapolation of $G_m(i\omega_n)$ to real
frequencies can be achieved by using, for example, the routine 
{\it ratint}\cite{ratint} based on an approximation in terms of 
rational functions. The polynomial basis of this extrapolation 
ensures that at real $\omega$ the correct continuous lattice 
spectrum is found for Im\,$G_m(\omega)$, rather than a discrete 
spectrum for fixed $n_s$, as obtained by fitting $G_{0,m}$ via 
the cluster impurity Green's function $G_{0,m}^{cl}$. We use 
the extrapolation routine {\it ratint} in the next section to 
evaluate the subband quasi-particle spectra $A_m(\omega)$ for 
Na$_{0.3}$CoO$_2$.  

\subsection{Central role of self-energy} 

We like to point out that, rather than back-transforming the
lattice Green's function from imaginary to real frequencies,
it is more appropriate to perform this transformation on 
the self-energy. In fact, the self-energy may be regarded as 
the central quantity determined by the DMFT self-consistent 
quantum-impurity treatment. Since ultimately one is interested
in the solid quasi-particle Green's function given in 
Eq.~(\ref{G}), the single-particle features determined by 
the Hamiltonian $H({\bf k})$ can evidently be evaluated at 
real $\omega$. Thus, only the real-$\omega$ variation of the
self-energy is needed. As a result, ${\bf k}$-resolved
spectra can be calculated via  
\begin{equation}
   A({\bf k},\omega)  = -\frac{1}{\pi}{\rm Im\, Tr}
     \Big(\omega + \mu - H({\bf k})  - \Sigma(\omega)
          \Big)^{-1} .
                                                   \label{Gk}
\end{equation}

The usefulness of transforming $\Sigma$ from Matsubara to 
real frequencies can be seen at small $U$. 
The spectral details of $G$ are then almost entirely of 
single-particle character, whereas the self-energy becomes 
very smooth. Since $\Sigma$ is a highly convoluted function, 
with contributions stemming from the occupied and unoccupied 
regions of the density of states, back-transforming it to 
real $\omega$ becomes progressively simpler at small $U$.
In contrast, the back-transformation of $G$ gets more
difficult. Obviously, since $H({\bf k})$ is known the
back-transformation of the single-particle part of $G$ 
can be avoided altogether. 

It is well known also that different cluster properties 
exhibit different convergence with increasing cluster size. 
Single-particle features, such as the local density of states
at the impurity in a given energy window, converge most slowly.
Integrated properties, such as the total energy, converge more 
rapidly. Since the phase space for two-particle interactions 
increases extremely fast with cluster size, many-body properties 
such as the self-energy exhibit even more favorable convergence
properties. On the other hand, the cluster Green's function
involves both single-electron and many-electron features. Only 
the latter converge rapidly with cluster size. These different   
behaviors also suggest that it is useful to regard 
$\Sigma\approx\Sigma^{cl}$ as the key quantity derived within 
ED/DMFT.

To extrapolate $\Sigma_m(i\omega_n)$ to real frequencies it 
is first necessary to remove the Hartree-Fock limit given by 
$\Sigma_m^{\rm HF}(i\omega_n)=\Sigma_m(i\omega_n\rightarrow\infty)$.
The remainder can then be transformed to real $\omega$ by using the 
same extrapolation routine {\it ratint} as for the derivation of 
$G_m(\omega)$.\cite{MIT} As in the latter case, this routine
ensures that the correct continuous lattice self-energy is found,
rather the discrete version appropriate for the cluster self-energy. 
Since back-transformation of single-particle features is avoided,
and quasi-particle features are calculated from Eq.~(\ref{G}) 
at real $\omega$, finer spectral details are retained than when
$G_m(i\omega_n)$ is transformed to the real axis. Nevertheless,  
in the following section we show that both schemes yield consistent 
spectra.

Before closing this section we note that the back-transformation 
of the self-energy from Matsubara to real frequencies should also 
be useful within QMC/DMFT. By viewing QMC as a tool for evaluating
$\Sigma$, the same separation of many-body features from 
single-particle aspects can be achieved if quasi-particle spectra
are evaluated from Eq.~(\ref{G}) at real $\omega$. Of course, the     
statistical uncertainties associated with QMC make the transformation 
of $\Sigma(i\omega_n)$ more subtle than in the present ED case.
Nevertheless, after subtraction of the Hartree-Fock limit from the 
real part of $\Sigma$, and after proper normalization (determined by 
the asymptotic behavior of the imaginary part), the use of the 
maximum entropy method on $\Sigma(i\omega_n),\,\Sigma(\tau)$ should 
be more accurate than on $G(i\omega_n),\,G(\tau)$, since the 
back-transformation of single-particle features is avoided.

\section{ Application to Na$_{0.3}$CoO$_2$ }

We have applied the multi-band ED/DMFT approach discussed in
the previous section to investigate the correlation induced
charge transfer between $t_{2g}$ bands of Na$_{0.3}$CoO$_2$.
The controversial small Fermi surface pockets arise from $e'_g$ 
bands that extend less than 100~meV above $E_F$. Since onsite 
Coulomb energies are much larger than the $t_{2g}$ band width,
it is important to inquire whether correlations can lead to a 
filling of the $e'_g$ bands. 

The single-particle properties for this Na concentration were 
derived from a full-potential linear augmented plane wave (LAPW) 
calculation, which was then fitted to a $3\times3$ $t_{2g}$
tight-binding Hamiltonian including three neighboring shells
and $dd\sigma$, $dd\pi$ and $dd\delta$ matrix elements. Further
details concerning the LDA band structure are given in 
Ref.~\cite{ishida}.

For the combined $t_{2g}$ and $e_g$ bands of Na$_{0.3}$CoO$_2$ 
\ $U$ was calculated to be about $3.7$~eV.\cite{johannes}
For the $t_{2g}$ subbands a smaller value might be appropriate
to account for screening via $e_g$ electrons. Since precise
values of $U$ and $J$ are not known, ED/DMFT calculations
up to $U=5$~eV were carried out, assuming $J=U/4$. 

\begin{figure}[t!]  
  \begin{center}
   \includegraphics[width=5.0cm,height=8cm,angle=-90]{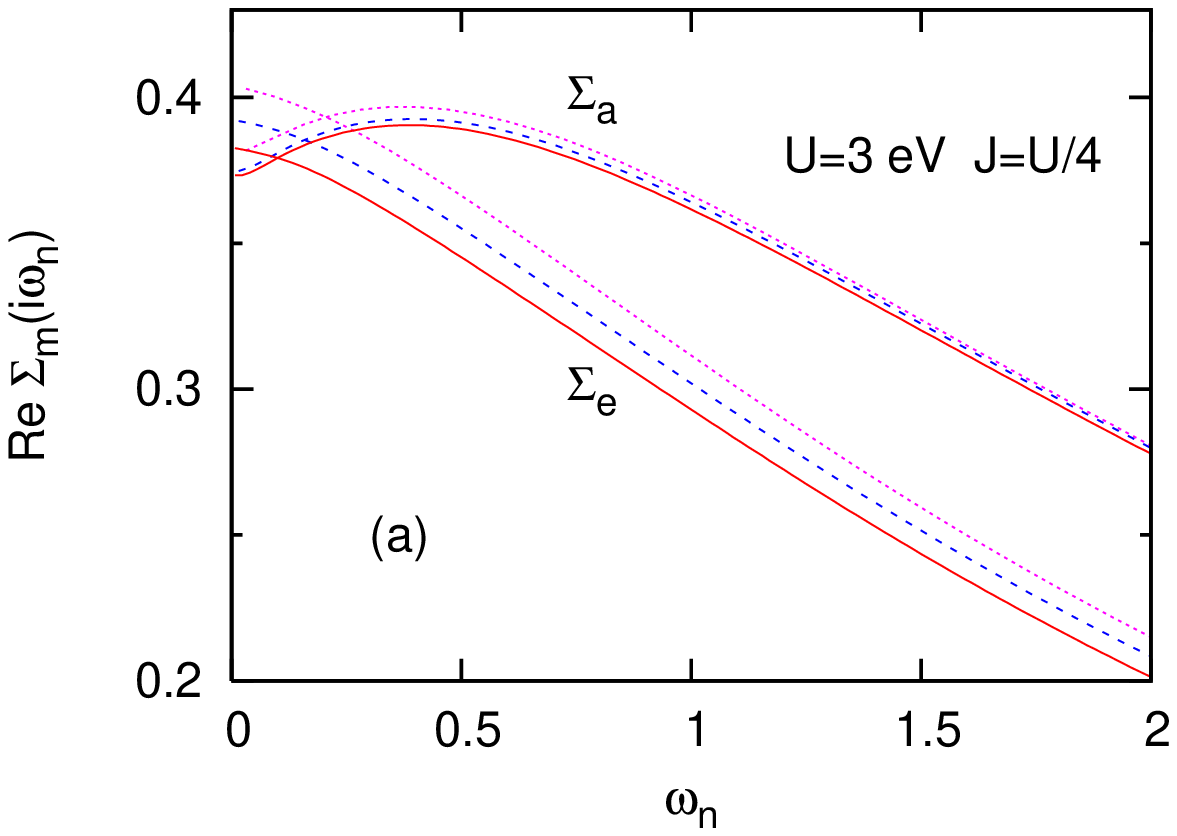}
   \includegraphics[width=5.0cm,height=8cm,angle=-90]{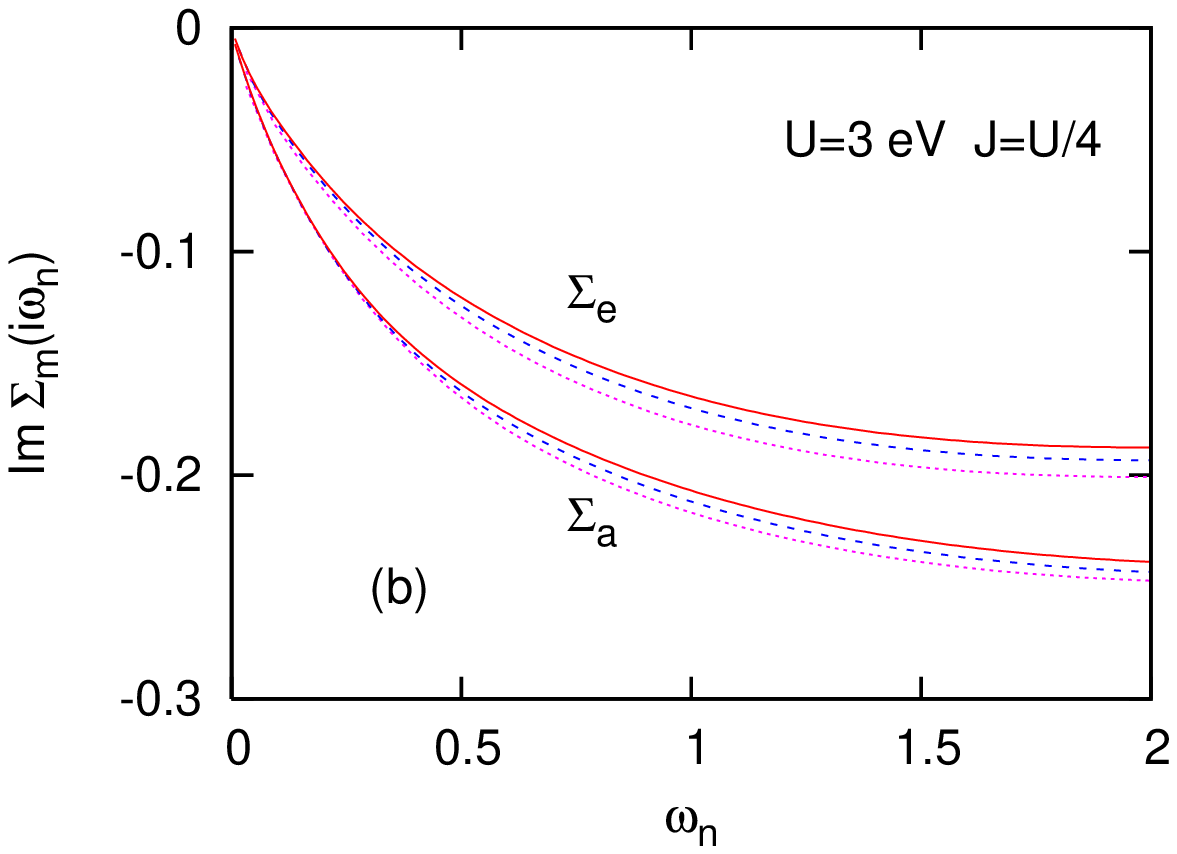}
  \end{center}
\caption{Subband self-energies $\Sigma_m(i\omega_n)$ of  
Na$_{0.3}$CoO$_2$ as calculated within ED/DMFT for $U=3.0$~eV, 
$J=0.75$~eV at several temperatures: red curves: $T=2.5$~meV, 
blue curves:  $T=5$~meV, magenta curves: $T=10$~meV. The 
Hartree-Fock limit is subtracted from Re\,$\Sigma_m(i\omega_n)$.
}\end{figure}

Fig.~1 shows the Co $a_g$ and $e'_g$ subband self-energies 
$\Sigma_m(i\omega_n)$ for $U=3$~eV and $J=0.75$~eV at 
$T=10$~meV. Hund exchange is fully included, i.e., $J'=J$. 
As a result of the large cluster size the spacing between 
excited states of the impurity Hamiltonian is less than 
$0.001$~eV. Although each impurity orbital hybridizes only 
with three bath levels, the Coulomb and exchange interactions
at the impurity site induce coupling between all bath levels, 
so that the excited states of the total cluster are significantly 
more closely spaced than for $n_s=4$ in the single-band case.  
Thus, even at temperatures as low as $T=2.5$~meV finite-size
effects are extremely small. The slight differences of the real 
part of the self-energy are related to the temperature dependence 
of the chemical potential.

The $a_g$ and $e'_g$  quasi-particle weights 
$Z_m = 1/[1- d{\rm Re}\Sigma_m( \omega  )/d\omega  ]\approx 
       1/[1-  {\rm Im}\Sigma_m(i\omega_0)/ \omega_0]$
derived from these self-energies are $Z_{a_g}\approx 0.52$
and $Z_{e'_g}\approx 0.62$, giving effective masses
$m^*_{a_g}\approx 1.9$ and $m^*_{e'_g}\approx 1.6$, respectively.  

\begin{figure}[t!]  
  \begin{center}
   \includegraphics[width=5.0cm,height=8cm,angle=-90]{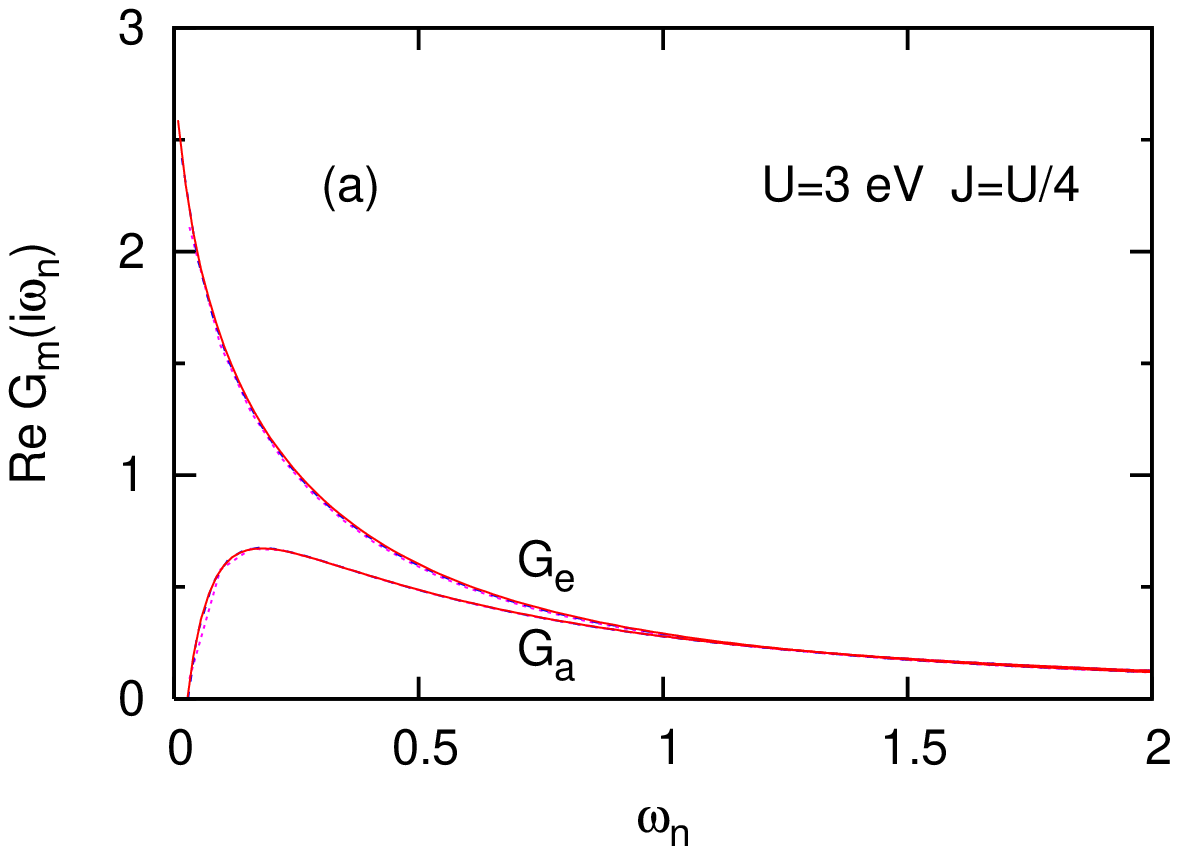}
   \includegraphics[width=5.0cm,height=8cm,angle=-90]{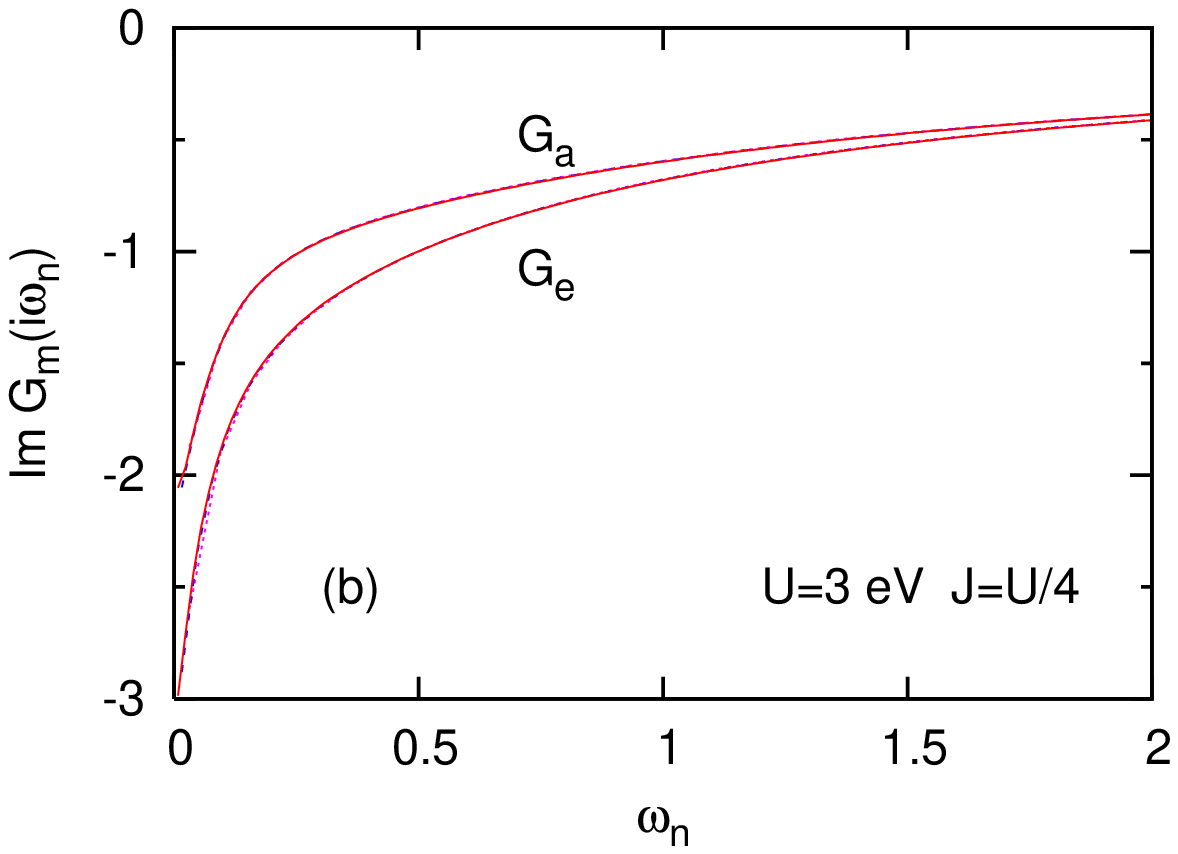}
  \end{center}
\caption{Subband Green's functions $G_m(i\omega_n)$ of  
Na$_{0.3}$CoO$_2$ for $U=3.0$~eV, $J=0.75$~eV, at several 
temperatures: red curves: $T=2.5$~meV, blue curves:  $T=5$~meV, 
magenta curves: $T=10$~meV. 
}\end{figure}

The corresponding subband lattice Green's functions are plotted 
in Fig.~2 for the same parameters as in Fig.~1. As in the case 
of the self-energies, the distributions are very smooth and 
exhibit only very small deviations caused by finite-size effects.
  
\begin{figure}[t!]  
  \begin{center}
   \includegraphics[width=5.0cm,height=8cm,angle=-90]{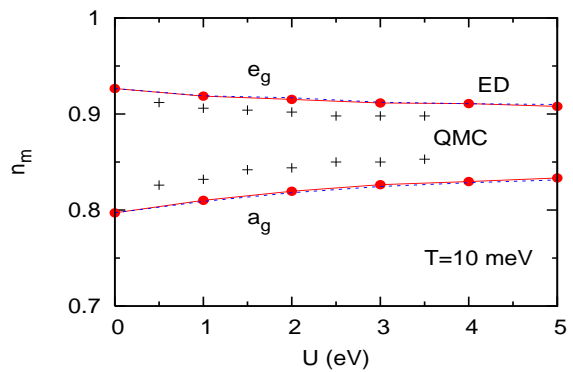}
  \end{center}
\caption{Subband occupations $n_{a_g}$ and $n_{e'_g}$ (per spin band) 
of Na$_{0.3}$CoO$_2$ as a function of $U$, with $J=U/4$ at $T=10$~meV. 
Solid (red) curves: Hund exchange $J'=J$, dashed (blue) curves:
Ising exchange $J'=J$. The corresponding QMC/DMFT results\cite{ishida} 
for $T=30\ldots60$~meV are indicated by the symbols (+).
The total occupation is $2n_{a_g}+4n_{e'_g}=5.3$.
}\end{figure}

Fig.~3 shows the variation of the ${a_g}$ and 
${e'_g}$ occupations with onsite Coulomb energy at $T=10$~meV. 
Both isotropic Hund exchange ($J'=J$) and Ising 
exchange ($J'=0$) are seen to give nearly identical charge 
transfer from ${e'_g}$ to ${a_g}$ bands. In the present ED/DMFT
treatment this transfer is slightly less pronounced than the 
one found within QMC/DMFT for Ising exchange at higher $T$, 
\cite{ishida} but the trends in both DMFT calculations are 
consistent. Thus, both DMFT treatments suggest that local
Coulomb correlations stabilize the ${e'_g}$ Fermi surface 
pockets. The overall topology of the Fermi surface therefore
remains the same as predicted by the LDA, i.e., correlation 
effects cannot explain the absence of the ${e'_g}$ pockets
from the ARPES data. 

Note that up to $U=5$~eV the subband occupations show no sign
of any reversal of charge transfer. The opposite result found
using the Gutzwiller approach\cite{gutzwiller} in the $U=\infty$
limit might therefore be a consequence of the approximate 
treatment of dynamical correlations. In this scheme the true 
complex frequency dependent self-energy is replaced by real 
coefficients representing the shift and narrowing of subbands.

\begin{figure}[t!]  
  \begin{center}
  \includegraphics[width=5.0cm,height=8cm,angle=-90]{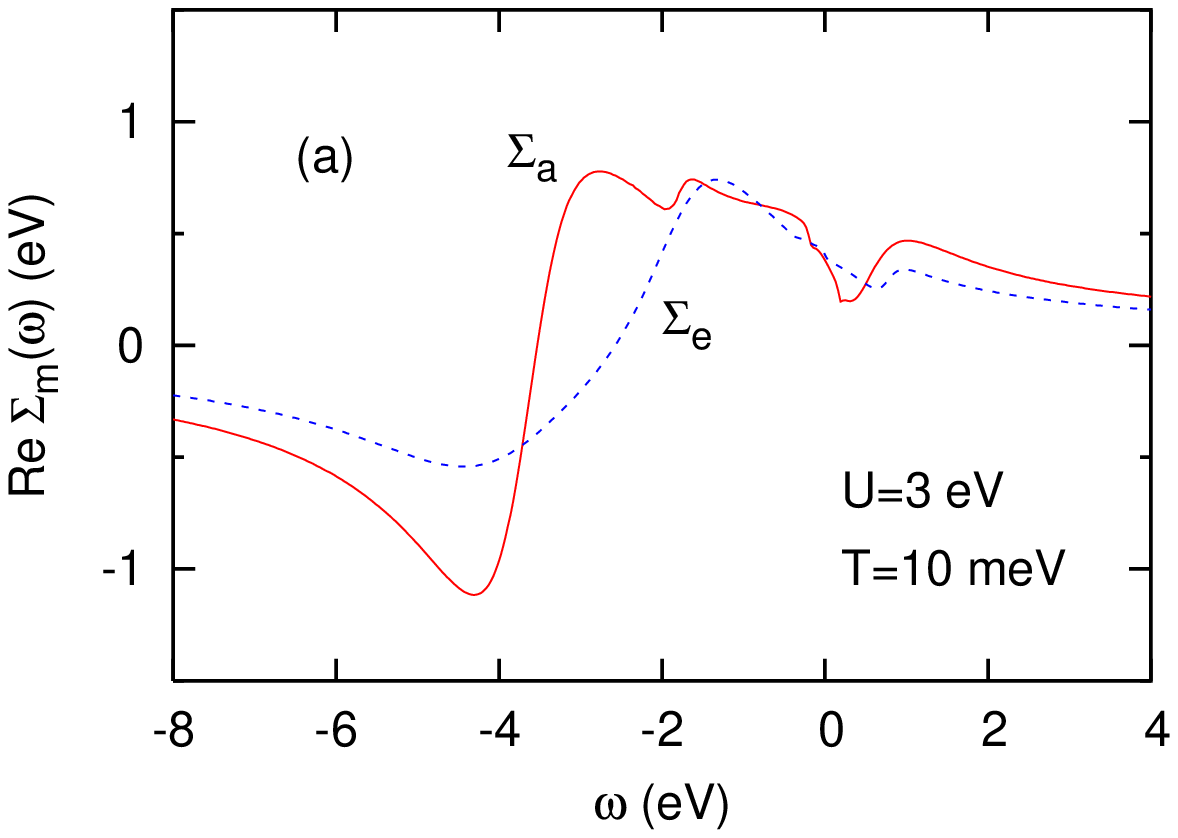}
  \includegraphics[width=5.0cm,height=8cm,angle=-90]{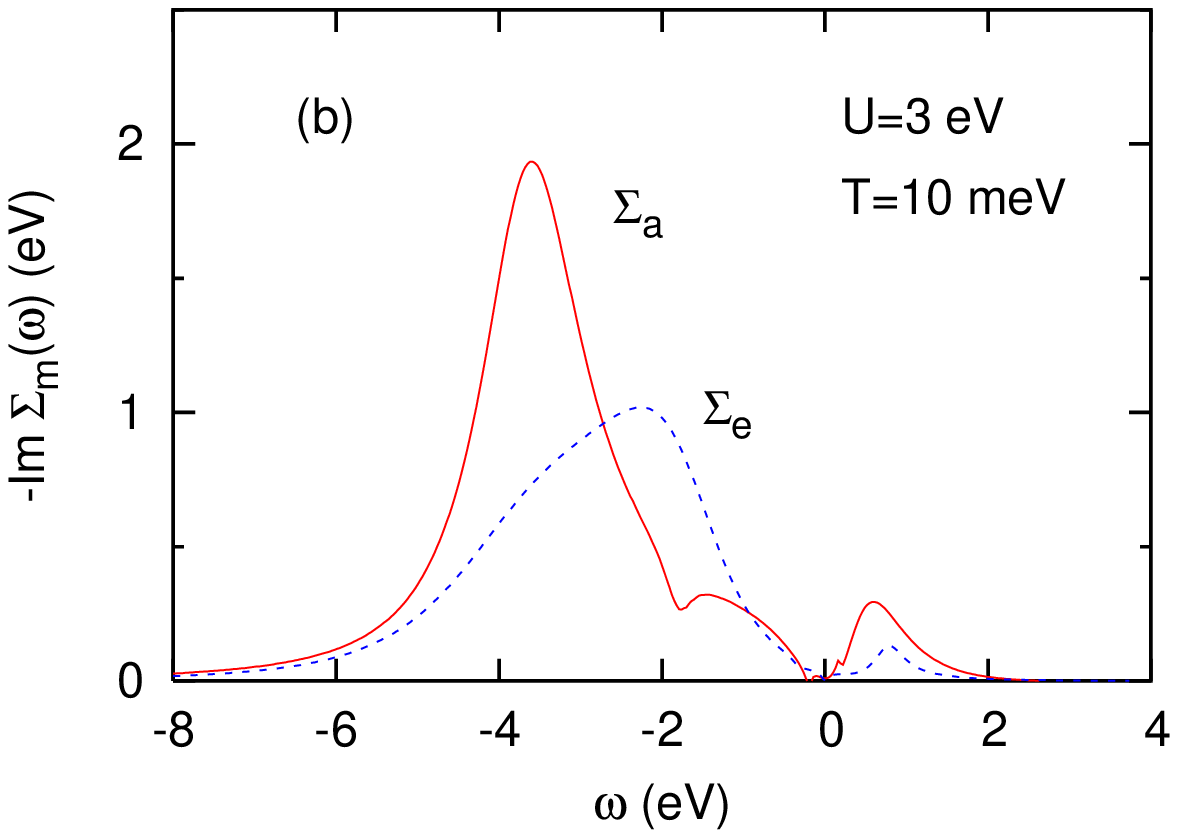}
  \end{center}
\caption{Subband self-energies $\Sigma_m(\omega)$ of Na$_{0.3}$CoO$_2$
at real frequencies, for the same parameters as in Fig.~1. Solid 
(red) curves:  ${a_g}$ bands, dashed (blue) curves: ${e'_g}$ bands. 
(a) Re\,$\Sigma_m(\omega)$; (b) $-$Im\,$\Sigma_m(\omega)$. 
The Hartree-Fock limit is subtracted from the real part.
}\end{figure}

Fig.~4 shows the frequency variation of the subband self-energies,
as obtained directly from $\Sigma_m(i\omega_n)$ via the  
extrapolation routine {\it ratint}. The Kramers-Kronig relations 
between the real and imaginary parts of $\Sigma_m(\omega)$ are 
very well satisfied. The striking asymmetry between the negative 
and positive frequency regions is a consequence of the near 
filling of the subbands.

\begin{figure}[t!]  
  \begin{center}
 \includegraphics[width=5.0cm,height=8cm,angle=-90]{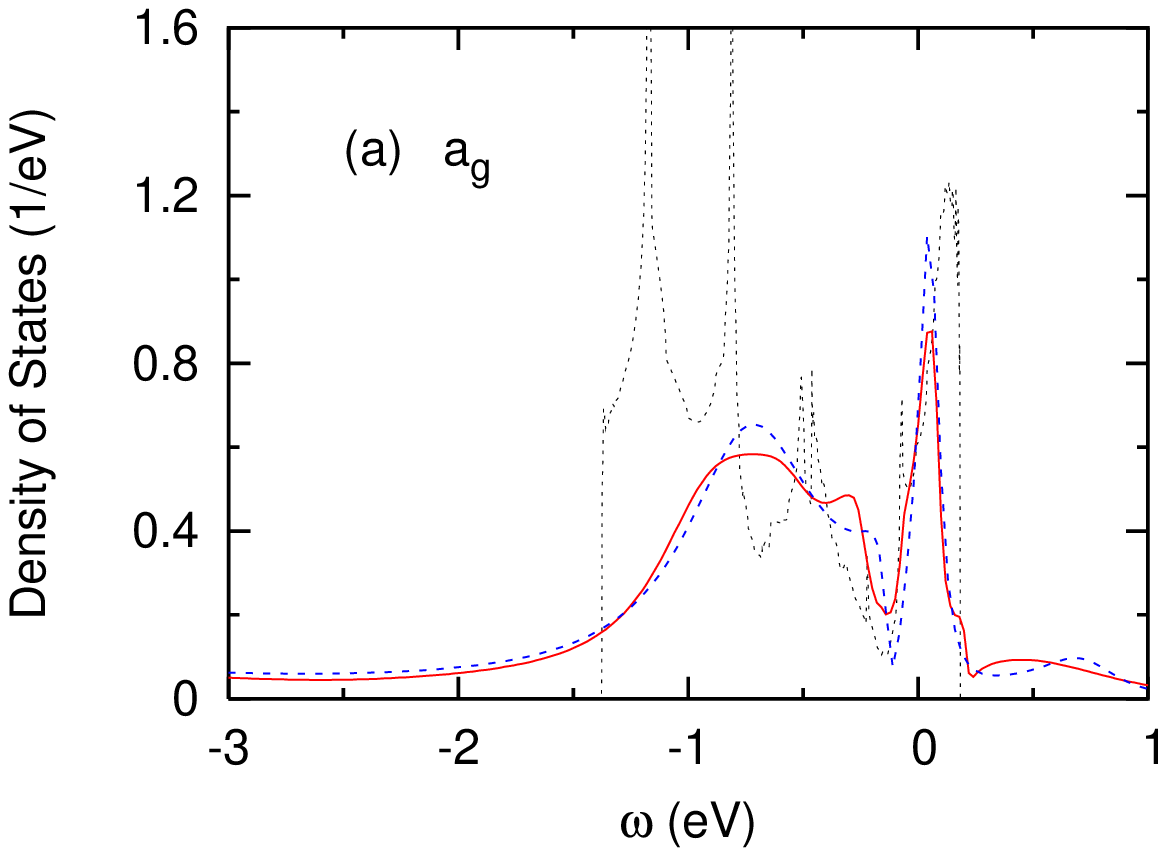}
 \includegraphics[width=5.0cm,height=8cm,angle=-90]{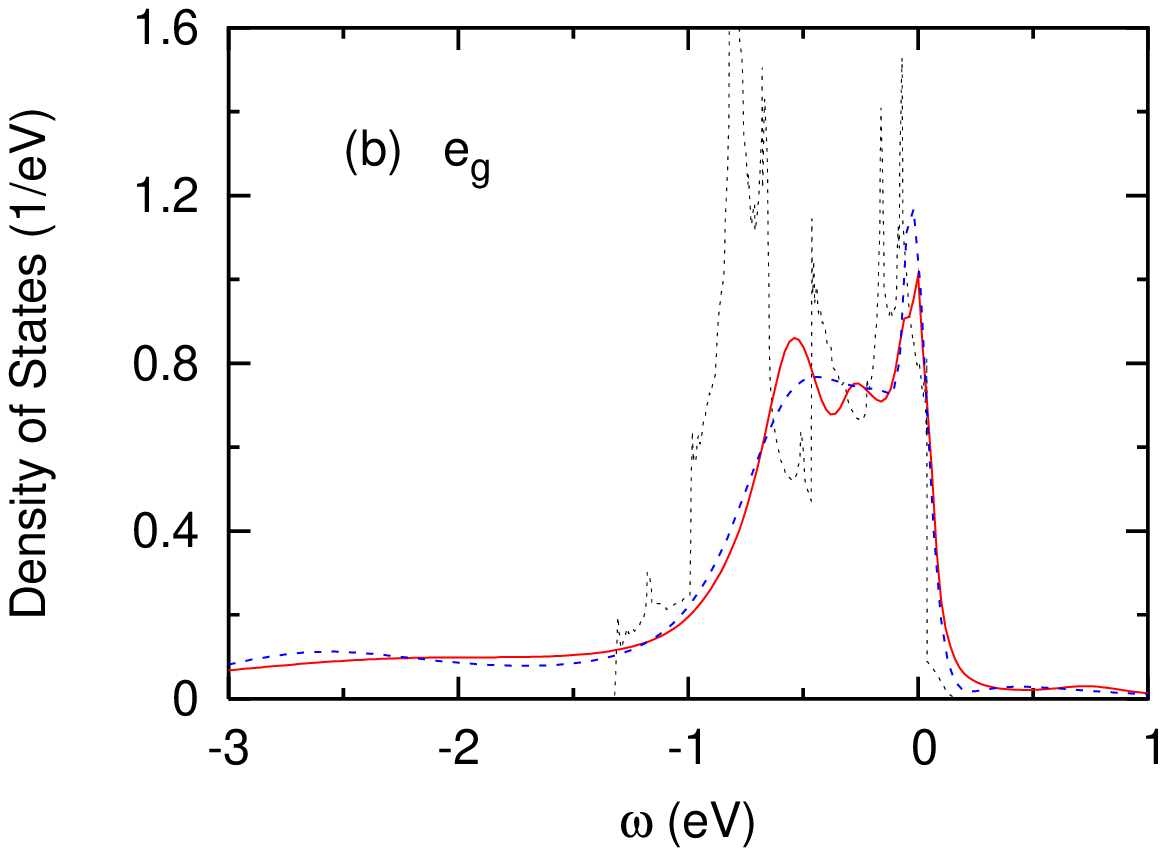}
  \end{center}
\caption{Quasi-particle spectra of (a) ${a_g}$ and (b) ${e'_g}$ 
subbands of Na$_{0.3}$CoO$_2$, calculated within ED/DMFT 
for the same parameters as in Fig.~1.
Solid (red) curves: spectra derived from $G_m(\omega)$ via 
$\Sigma_m(\omega)$;
dashed (blue) curves: spectra obtained by extrapolating 
$G_m(i\omega_n)$ to real $\omega$;
dotted (black) curves: single-particle density of states.
}\end{figure}

As pointed out in the previous section, quasi-particle spectra
at real frequencies can be derived by back-transforming the solid
Green's function $G_m(i\omega_n)$, or by first transforming 
$\Sigma_m(i\omega_n)$ and then applying Eq.~(\ref{G}) at real 
$\omega$. The comparison shown in Fig.~4 proves that both methods
are consistent, and that the latter scheme retaines finer
spectral details originating from the single-particle 
Hamiltonian. For instance, the ${e'_g}$ spectrum obtained 
via Eq.~(\ref{G}) and $\Sigma_m(\omega)$ shows two peaks
below $E_F$ which evidently are the shifted and broadened
density of states features near 0.4 and 0.8~eV below the
Fermi level. Also, the peak close to $E_F$ exhibits some
of the fine structure of the single-particle density of
states. These details are lost if the spectrum is instead
derived via back-transformation of $G_m(i\omega_n)$.

Interestingly, the ${a_g}$  spectrum has a steep positive 
slope at $E_F$, while the ${e'_g}$ spectrum has a strong
negative slope. This holds true for both methods of 
evaluating the real frequency spectra. Also, the ${a_g}$
quasi-particle spectrum exhibits a pronounced minimum at 
about $0.1$~eV below $E_F$. This feature exists already 
in the single-particle density of states and is evidently
not obliterated by correlation effects.

Both subbands exhibit the usual correlation induced band 
narrowing, quasi-particle damping, and incoherent spectral 
weight in the region below the bottom of the bands,
associated with Hubbard peaks. Qualitatively, these spectra
agree with the ones derived previously using the QMC/DMFT
and the maximum entropy method.\cite{ishida,lechermann,marianetti} 

\begin{figure}[t!]  
  \begin{center}
 \includegraphics[width=5.0cm,height=8cm,angle=-90]{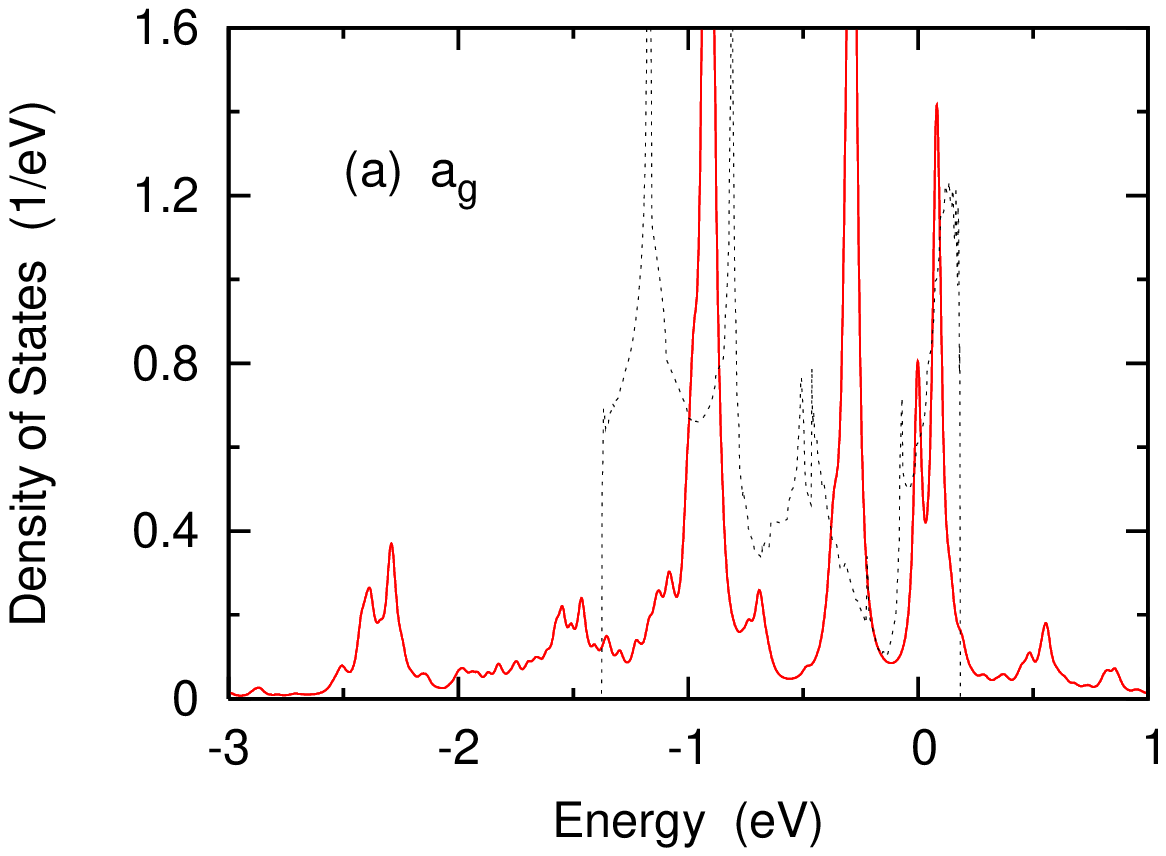}
 \includegraphics[width=5.0cm,height=8cm,angle=-90]{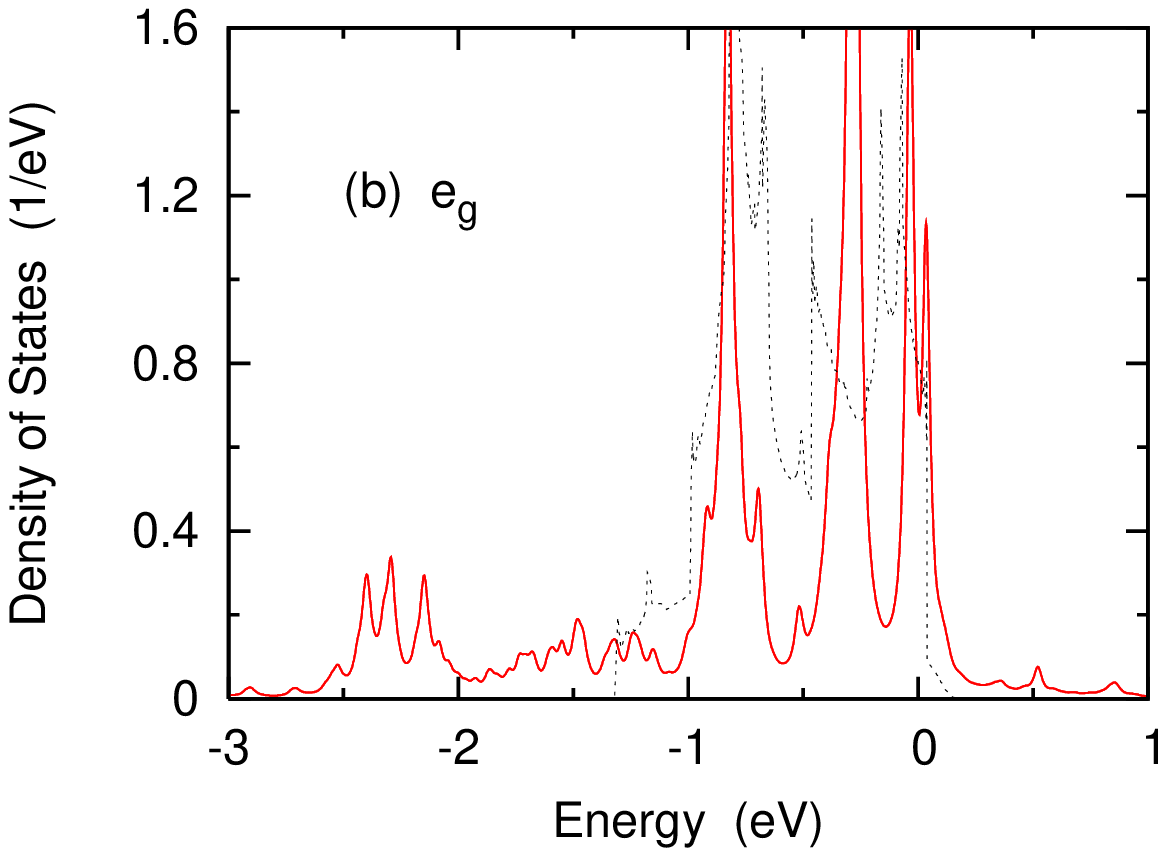}
  \end{center}
\caption{Cluster spectra of (a) ${a_g}$ and (b) ${e'_g}$ 
orbitals of Na$_{0.3}$CoO$_2$, as calculated from Eq.~(\ref{Gcl}).
Solid (red) curves: $-$Im\,$G_m^{cl}(\omega+i\delta)$ with
$\delta=25$~meV; 
dotted (black) curves: single-particle density of states.
}\end{figure}

For comparison with these lattice spectra we show in Fig.~5
the cluster spectra obtained by evaluating Eq.~(\ref{Gcl}) at 
$\omega+i\delta$ instead of $i\omega_n$. Because of the large 
cluster size, the level spacing between excited states in the 
metallic phase of Na$_{0.3}$CoO$_2$ is less than $10^{-3}$~eV. 
Thus, with only a minor broadening 
these cluster spectra look very smooth indeed. The spectral
weight is located mainly in the $t_{2g}$ band region, and 
there is clear evidence of incoherent weight associated with 
Hubbard bands. Both the ${a_g}$ and ${e'_g}$ spectral 
distributions exhibit three main features in the band region,
as in the case of the bare density of states. Nevertheless, 
these features are much narrower than in the corresponding
lattice spectra shown in Fig.~4. This is not surprising since, 
as pointed out in the previous section, the cluster Green's 
function depends on single-particle and many-body aspects. 
Only the latter, represented by the self-energy, converge rapidly 
with cluster size. Thus, increasing the artificial broadening of
the cluster spectra does not yield the lattice spectra.
Cluster spectra would resemble more closely those of the solid 
only if $n_s$ is significantly increased. To analize photoemission 
spectra of the solid material, Eq.~(\ref{G}) should therefore 
be used at real $\omega$ rather than Eq.~(\ref{Gcl}).

Transforming the subband self-energies $\Sigma_{a_g}(\omega)$
and $\Sigma_{e'_g}(\omega)$ to the $t_{2g}$ basis, they can 
be used to evaluate the momentum dependence of the
quasi-particle band structure, as indicated in Eq.~(\ref{Gk}).
This is shown in Fig.~7 for the $\rm \Gamma K$ and $\rm \Gamma M$
symmetry directions of the hexagonal Brillouin Zone. The
 ${e'_g}$ Fermi surface pockets are associated with the flat
bands reaching just above $E_F$ along $\rm \Gamma K$.
The maximum of this band is barely shifted, but the crossing of 
$E_F$ occurs slightly farther away from the maximum, leading to 
a weak enlargement of the hole pocket. Conversely, the diameter 
of the large $a_g$ pocket centered at $\rm\Gamma$ is slightly 
reduced.      

\begin{figure}[t!]  
  \begin{center}
  \includegraphics[width=8.5cm,height=7cm,angle=0]{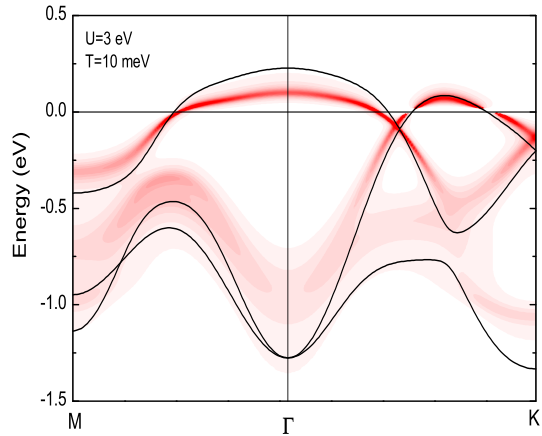}
  \end{center}
  \vskip-2mm
\caption{Dispersion of quasi-particle bands of Na$_{0.3}$CoO$_2$ 
along $\rm M\Gamma$ and $\rm \Gamma K$ for $T=10$~meV. 
The LDA bands derived from $H({\bf k})$ are denoted by the 
solid lines.
}\end{figure}

These results demonstrate that, although correlation effects 
induce band narrowing and broadening, they do not alter the basic 
dispersion of the $t_{2g}$ bands of Na$_{0.3}$CoO$_2$. In particular,
it does not seem possible to interprete these bands in terms of 
a single-band model, with a flat $a_g$ band limited to within 
a very narrow energy range near $E_F$.\cite{merino,qian}

\section{Summary and Outlook}

We have applied ED/DMFT to investigate the correlation induced 
charge transfer between the $t_{2g}$ bands in Na$_{0.3}$CoO$_2$.  
Since previous QMC/DMFT studies of this problem were limited 
to Ising exchange the role of full Hund's coupling had remained 
unresolved. The ED calculations show that the charge transfer  
follows the same trend for both Hund and Ising exchange: The 
$e'_g$ bands donate some of their charge to the $a_g$ bands,
i.e., the $e'_g$ Fermi surface hole pockets are slightly enlarged.
These ED results fully confirm the trend obtained previously
within QMC/DMFT for Ising exchange. Thus, local Coulomb 
correlations cannot explain the absence of 
these pockets from the ARPES data.   

These results suggest that ED/DMFT is a useful method for the study of 
strongly correlated three-band systems. By exploiting the sparseness 
of the impurity Hamiltonian the cluster size $n_s=12$ can now be 
investigated, without loss of accuracy, with about the same 
computational effort as $n_s=8$ using full diagonalization. 
The large cluster size ensures that the level spacing among
excited states is very small, so that finite-size effects are
greatly reduced.

In view of this improvement, ED/DMFT can now be regarded
as complementary to QMC/DMFT. We point out, however, that
in comparison to standard QMC/DMFT based on the Hirsch-Fye 
algorithm, the ED/DMFT approach is free of sign problems. 
Thus, complete Hund exchange and large on-site Coulomb 
interactions can be treated. In addition, very low temperatures 
can be reached. In principle, results for $T\rightarrow0$ can 
also be generated within ED/DMFT. This limiting region, however, 
requires special care since certain system properties might be 
sensitive to remaining finite-size effects even for $n_s=12$. 

We have also demonstrated that ED/DMFT provides continuous 
quasi-particle spectra and self-energies of the extended solid 
at real frequencies. Moreover, by considering the self-energy 
as the principal quantity derived within DMFT, a clear separation 
of single-electron and many-body aspects in spectral distributions 
can be achieved.    

In the future it should be valuable to apply ED/DMFT to other
multi-band materials that have so far been studied only within
Ising exchange and QMC/DMFT, but where full Hund's coupling is 
crucial. Also, by exploiting the fact that 
the present scheme can easily be parallelized, much larger cluster 
sizes are feasible. This opens the path for the study of various 
other material properties, including non-local effects.

\bigskip

{\bf Acknowledgements:} We like to thank Eric Koch for drawing 
our attention to the Arnoldi algorithm. We also thank him and 
Theo Costi for stimulating discussions.   
Most of the ED/DMFT calculations were carried out on the IBM 
supercomputer (JUMP) of the Forschungszentrum J\"ulich.

\bigskip

E-mail: a.perroni@fz-juelich.de;  
         ishida@chs.nihon-u.ac.jp; 
         a.liebsch@fz-juelich.de


\begin{thebibliography}{99}
\bibitem{MIT}
   We focus here on the metallic phase. Near a metal 
   insulator transition, the singular part of the 
   self-energy has to be treated separately.
\end{thebibliography}
\end{document}